\documentclass[final]{aipproc}
\pdfoutput=1

\layoutstyle{6x9}

\usepackage{graphicx,amsmath,amssymb}
\usepackage{subfig}

\def \d {{\rm d}}

\def \l {\ell}

\begin{document}

\title{Conformal infinity in Robinson--Trautman spacetimes with cosmological constant}
\author{Otakar Sv\'{\i}tek}{address={Institute of Theoretical Physics, Charles University, Prague}}

\begin{abstract}
In past, the future asymptotic behavior (with respect to initial data on null hypersurface) of Robinson--Trautman spacetime was examined and its past horizon characterized. Therefore, only the investigation of conformal infinity is missing from the picture. We would like to present some initial results concerning conformal infinity when negative cosmological constant is present motivated by the AdS/CFT correspondence.
\end{abstract}

\maketitle

\section{Introduction}
Various aspects of an important Robinson--Trautman family of expanding, shearfree and twistfree spacetimes \cite{RobinsonTrautman:1960,RobinsonTrautman:1962} were studied during the last 50 years. The existence, asymptotic behavior, possible extensions beyond future horizon and other specific properties of \emph{cosmological and/or pure radiation} solutions of algebraic type~II with  spherical topology were investigated, see \cite{Chru1,Chru2,ChruSin,podbic95,podbic97,podsvit05,rezzolla} for more recent expositions. Also the past horizon properties were extensively covered \cite{ChowLun:1999,NatorfTafel:2008,PodolskySvitek:2009}. Several results were also generalized to higher dimensional version of this family \cite{podolsky-ortaggio,Svitek:2011}.

Having global causal structure in mind, one is naturally interested in learning something about conformal infinity. Here, we will show initial results concerning Robinson--Trautman spacetimes with negative cosmological constant.

Inspired by the fact that the selected subfamily is locally asymptotically AdS at timelike conformal infinity we will use the view provided by AdS/CFT correspondence \cite{Maldacena,Witten}. Namely, we will determine the associated boundary stress energy tensor with the possible identification of dual field theory in mind. For black hole spacetimes the favored interpretation is in terms of strongly coupled fluid \cite{Hubeny}.


Robinson--Trautman spacetimes of Petrov type~II (with a cosmological constant $\Lambda$ and pure radiation with density $w$) can be described by the following metric
\begin{equation}\label{RT-metric}
\d s^2 = -2H\,\d u^2-\,2\,\d u\,\d r +2\,\frac{r^2}{P^2}\,\d\zeta\,\d\bar\zeta
\end{equation}
where  ${2H = \Delta(\,\ln P) -2r(\,\ln P)_{,u} -{2m(u)/r} -(\Lambda/3) r^2}$. Using the Laplacian $\Delta\equiv 2P^2\partial_{\zeta}\partial_{\bar\zeta}$ the field equations reduce to the well-known Robinson--Trautman equation
\begin{equation}
\Delta\Delta(\,\ln P)+12\,m(\,\ln P)_{,u}-4\,m_{,u}=2\kappa\ w^2
\end{equation}

\begin{figure}
\captionsetup{type=figure}
\centering{
 \subfloat{\includegraphics[width=0.4\textwidth]{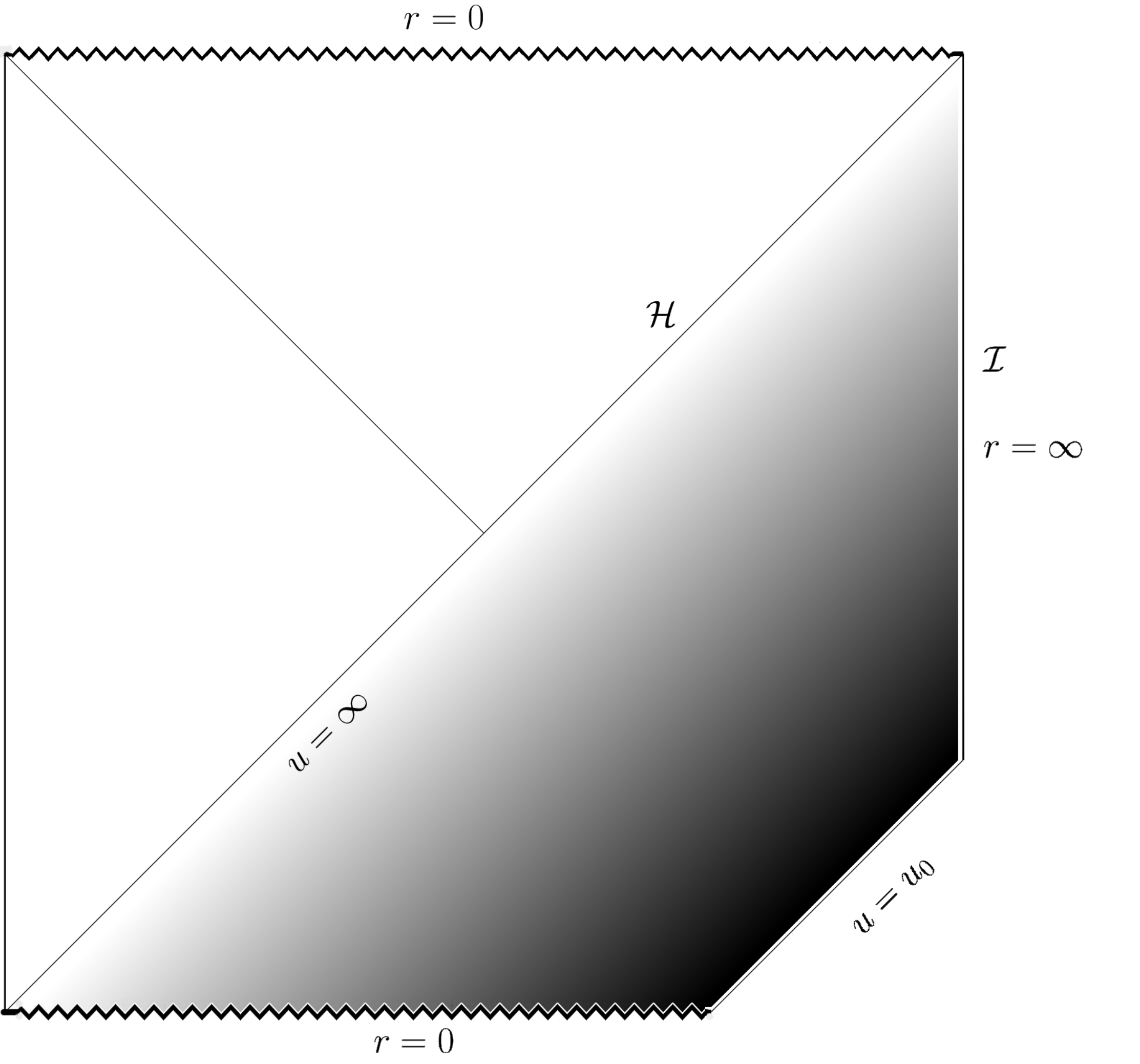}}
\hfill
 \subfloat{\includegraphics[width=0.4\textwidth]{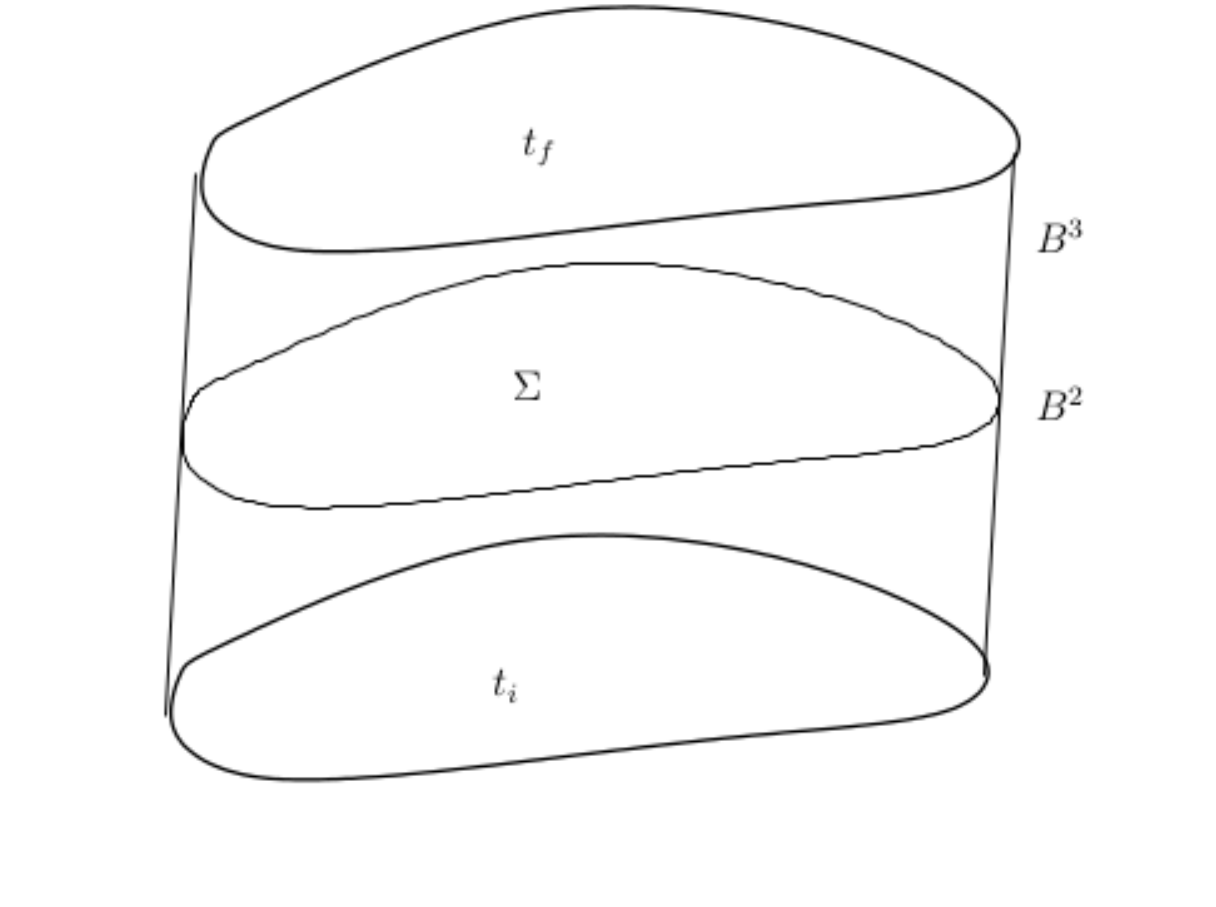}}
\caption{\hspace{-0.2cm}(a) Robinson--Trautman-AdS \hspace{1.8cm}  (b) Foliation}}
\end{figure}

These spacetimes have well defined evolution into the future of a null hypersurface ($u=u_{0}$, see Figure 1a) with arbitrary smooth initial data. Asymptotically (as $u\to\infty$) they approach Schwarzschild--((anti-) de Sitter) or Vaidya--((anti-) de Sitter) when pure radiation is present. Past black hole horizon can be localized quasilocally using any of the following definitions: apparent horizon, trapping horizon, dynamical horizon.

\section{Boundary stress energy tensor for asymptotically AdS spacetimes}
We will try to compute the boundary stress energy tensor (SET) associated with the gravitating asymptotically AdS spacetime. The result can be generally interpreted as an expectation value of the stress energy tensor in a dual conformal field theory (AdS/CFT).

To compute the SET we will use a method developed by J.D. Brown and J.W. York \cite{Brown-York}(so called Brown-York SET). The derivation is based on Hamilton-Jacobi theory. They consider a spacetime region $M=\Sigma \times R$ with metric $g$ and denote $B^{2}$ the two-dimensional boundary of $\Sigma$ (equipped with metric $h$). Further, $B^{3}$ (with metric $\gamma$) is the timelike part of $\partial M$ which is sliced by two-dimensional surfaces $B^{2}$ (see Figure 1b).

Now, consider the following gravitational action with negative cosmological constant, boundary terms and matter
\begin{equation}
S=\frac{1}{2\kappa}\int_{M}d^{4}x\sqrt{-g} \left(R+\frac{6}{\l^2}\right) + \frac{1}{\kappa}\int_{t_{i}}^{t_{f}}d^{3}x\sqrt{h}\,K - \frac{1}{\kappa}\int_{B^{3}}\sqrt{-\gamma}\ \Theta + S_{\mathrm{matter}}
\end{equation}
where $\Lambda=-\frac{3}{\l^2}$. In a standard way, arbitrary variation gives terms corresponding to equations of motion and relevant boundary terms. However, when the action $S$ is evaluated at classical solution ($S$ $\to$ $S_{\mathrm{cl}}$) it is a functional of boundary data: $\gamma,h(t_{i}), h(t_{f})$. So restricting only to variations among classical solutions we determine the dependence of $S_{\mathrm{cl}}$ on them (omitting indices for now)
\begin{equation}
\delta S_{\mathrm{cl}}= \{\mathrm{variations\ of\ matter\ fields}\} + \int_{t_{i}}^{t_{f}}d^{3}x\ P_{\mathrm{cl}}\,\delta h + \int_{B^{3}}d^{3}x\ \pi_{\mathrm{cl}}\,\delta\gamma
\end{equation}
where $\pi_{\mathrm{cl}}$ is the gravitational momentum conjugate to $\gamma$. Quasilocal boundary SET is then defined in analogy with mechanical systems as
\begin{equation}\label{BY-SET}
T=\frac{2}{\sqrt{-\gamma}}\frac{\delta S_{\mathrm{cl}}}{\delta\gamma}=\frac{2}{\sqrt{-\gamma}}\pi_{\mathrm{cl}}
\end{equation}
However, the prescription diverges as the boundary is taken to infinity and therefore some form of regularization is needed. We will select the approach by Balasubramanian and Kraus \cite{Balasubramanian} who proposed a renormalization procedure consisting of subtracting suitable counterterms. These counterterms consist only from boundary terms and so they do not influence bulk equations of motion.

To arrive at the formula we start with ADM-like decomposition (foliation is timelike)
\begin{equation}
\d s^2=N^2\d r^2 + \gamma_{\mu\nu}(\d x^{\mu}+N^{\mu}\d r)(\d x^{\nu}+N^{\nu}\d r)
\end{equation}
Since the boundary metric $\gamma$ acquires an infinite Weyl factor asymptotically we will formally consider conformal class of boundaries. We assume the action (from now we are only interested in dependence on boundary metric $\gamma$) in the form
$S_{\gamma}=S_{\mathrm{cl}}[\gamma]+\frac{1}{\kappa}S_{\mathrm{ct}}[\gamma] $
and introduce the extrinsic curvature $\Theta^{\mu\nu}=-\nabla^{(\mu}n^{\nu)}$, where $\mathbf{n}$ is an outward pointing normal to $B^{3}$.

Then, $S_{\mathrm{ct}}$ is constructed by requiring cancellation of all divergences using technique resembling tree unitarity approach in QFT. Specifically, using prescription (\ref{BY-SET}) with $S_{\gamma}$ in four dimensions \cite{Balasubramanian} arrive at
\begin{equation}\label{SET}
T^{\mu\nu}=\frac{1}{\kappa}\left[\Theta^{\mu\nu}-\Theta\,\gamma^{\mu\nu}-\frac{2}{\l}\gamma^{\mu\nu}-\l\,G^{\mu\nu}(\gamma)\right]
\end{equation}
Taking an additional ADM foliation (now spacelike) of boundary one can compute interesting physical quantities by projecting $T^{\mu\nu}$ on the vector generating time flow on $B^{3}$.

\section{Robinson--Trautman-AdS boundary SET}
After introducing new coordinate $l=r^{-1}$ and taking the conformal factor to be ${\Omega=l}$, the Robinson--Trautman metric becomes
$
\d s^2=\Omega^{-2}[({\textstyle\frac{1}{3}}\Lambda +2\,l(\,\ln P)_{,u} -l^2 \Delta\ln P+ 2m\,l^3 ) \d u^2 + 2\,\d u\,\d l + 2\,P^{-2}\,\d\zeta\,\d\bar\zeta\,]
$
so it is locally asymptotically AdS for negative cosmological constant.

Before applying the previously derived results we have to account for the additional boundary segment corresponding to hypersurface $u=u_{0}$. Since this boundary is equipped with initial data determining the bulk evolution we have to assume that its boundary metric is subject to variations corresponding to changes in the bulk solution. However, we will not investigate canonical pair for this boundary explicitly because it is not influencing the calculation of SET for conformal infinity.

Using a simple transformation $u=t-r$ on the metric (\ref{RT-metric}) we arrive at 
\begin{equation}
\d s^2=-2H\d t^2-2(1-2H)\,\d t \d r+2(1-H)\,\d r^2+\frac{r^2}{P^2}\,\d\zeta\,\d\bar\zeta
\end{equation}
which is suitable for reading of the necessary quantities:
$
\gamma_{\mu\nu}=\mathrm{diag}(-2H,{\textstyle\frac{r^2}{P^2}},{\textstyle\frac{r^2}{P^2}}),\, N=(2H)^{-1/2},\, N^{\mu}=(N^{2}-1,0,0),\, \mathbf{n}=N\d r
$.

From (\ref{SET}) we compute that the leading term of SET is
$T_{00}=\frac{2m(u)}{\l r}+\ \cdots$
which generalizes the previous results for Schwarzschild--AdS spacetime. Using the above mentioned additional ADM splitting on the boundary with parameters
$
\sigma_{ab}=\mathrm{diag}({\textstyle\frac{r^2}{P^2},\frac{r^2}{P^2}}),\ N_{B^{2}}=\sqrt{2H},\ \mathbf{v}=\sqrt{2H}\d t
$
we can compute the proper energy density on the boundary
\begin{equation}
\int_{B^{2}}d\zeta d\bar\zeta\, \sqrt{\sigma}N_{B^{2}}T_{\mu\nu}v^{\mu}v^{\nu}=\int_{B^{2}}d\zeta d\bar\zeta\, \frac{4m(u)^2}{P^2\l}
\end{equation}
which however differs from corresponding expression given by Robinson and Trautman that involved term $m^{2/3}$ instead of $m^2$.

\section{Conclusions and outlook}
The presented results are so far only preliminary and have to be analyzed and extended. Specifically we would like to understand the correspondence with strongly coupled fluid in this context. Having the causal structure in mind one is also interested in explicitly connecting the behavior at conformal infinity with the dynamics of past horizon.

\begin{theacknowledgments}
This work was supported by grant GACR 202/09/0772.
\end{theacknowledgments}


\begin{thebibliography}{20}

\bibitem{RobinsonTrautman:1960} I.~Robinson and A.~Trautman,
\emph{Phys. Rev. Lett.} {\bf 4}, 431 (1960).

\bibitem{RobinsonTrautman:1962} I.~Robinson and A.~Trautman,
\emph{Proc. Roy. Soc. Lond.} {\bf A265}, 463 (1962).

\bibitem{Chru1}   P.~T.~Chru\'{s}ciel,
\emph{Commun. Math. Phys.} {\bf 137}, 289 (1991).

\bibitem{Chru2}   P.~T.~Chru\'{s}ciel,
\emph{Proc. Roy. Soc. Lond.} {\bf A436}, 299 (1992).

\bibitem{ChruSin}  P.~T.~Chru\'{s}ciel and D.~B.~Singleton,
\emph{Commun. Math. Phys.} {\bf 147}, 137 (1992).

\bibitem{podbic95} J.~Bi\v{c}\'{a}k and J.~Podolsk\'{y},
\emph{Phys. Rev. D} {\bf 52}, 887 (1995).

\bibitem{podbic97} J.~Bi\v{c}\'{a}k and J.~Podolsk\'{y},
\emph{Phys. Rev. D} {\bf 55}, 1985 (1997).

\bibitem{podsvit05} J.~Podolsk\'{y} and O.~Sv\'{\i}tek,
\emph{Phys. Rev. D} {\bf 71}, 124001 (2005).

\bibitem{rezzolla} L. Rezzolla, R. P. Macedo and J. L. Jaramillo, \emph{Phys. Rev. Lett.} {\bf 104}, 221101 (2010).

\bibitem{ChowLun:1999}  E.~W.~M.~Chow and A.~W.~C.~Lun,
\emph{J. Austr. Math. Soc. B} {\bf 41}, 217 (1999).

\bibitem{NatorfTafel:2008}  W.~Natorf and J.~Tafel,
\emph{Class. Quantum Grav.} {\bf 25}, 195012 (2008).

\bibitem{PodolskySvitek:2009} J. Podolsk\'y and O. Sv\'{\i}tek, \emph{Phys. Rev. D} {\bf 80}, 124042 (2009).

\bibitem{podolsky-ortaggio} J. Podolsk{\'y} and M. Ortaggio, \emph{Class. Quant. Grav.} {\bf 23}, 5785 (2006). 

\bibitem{Svitek:2011} O. Sv\'{\i}tek, \emph{Phys. Rev. D} {\bf 84}, 044027 (2011).

\bibitem{Maldacena} J. M. Maldacena, \emph{Adv. Theor. Math. Phys.} {\bf  2}, 231 (1998).

\bibitem{Witten} E. Witten, \emph{Adv. Theor. Math. Phys.} {\bf  2}, 253 (1998).

\bibitem{Hubeny} S. Bhattacharyya, V. E. Hubeny, S. Minwalla and M. Rangamani, \emph{JHEP} {\bf 0802}, 045 (2008).

\bibitem{Brown-York} J. D. Brown and J. W. York, \emph{Phys. Rev. D} {\bf 47}, 1407 (1993). 

\bibitem{Balasubramanian} V. Balasubramanian and P. Kraus, \emph{Commun. Math. Phys.} {\bf 208}, 413 (1999)

\end{thebibliography}
\end{document}